\documentclass[5p,number,sort&compress]{elsarticle}
\usepackage[intlimits]{amsmath}
\usepackage{amssymb}
\usepackage{bm,mathrsfs}
\usepackage[mathcal]{euscript}
\usepackage[breaklinks=true,unicode=true,urlcolor = blue,colorlinks = true,citecolor = blue,linkcolor = blue]{hyperref}
\usepackage{graphicx}
\usepackage[subrefformat=parens]{subcaption}
\usepackage{color}
\usepackage[normalem]{ulem}
\usepackage{multirow}

\renewcommand{\vec}[1]{\bm{#1}}
\graphicspath{{./}}
\usepackage{pgf}
\usepackage{tikz}
\usetikzlibrary{arrows,shapes,backgrounds}
\usetikzlibrary{positioning,arrows}
\usetikzlibrary{calc}

\begin{document}

\title{Effects of Surface Anisotropy on Magnetic Vortex Core}

\author[cor1,knu]{Oleksandr V. Pylypovskyi}
\ead{engraver@univ.net.ua}

\author[knu]{Denis D. Sheka}

\author[bitp]{Volodymyr P. Kravchuk}

\author[bitp]{Yuri Gaididei}

\address[knu]{Taras Shevchenko National University of Kiev, 01601 Kiev, Ukraine}
\address[bitp]{Institute for Theoretical Physics, 03143 Kiev, Ukraine}

\date{November 27, 2013}

%
%

\begin{abstract}
The vortex core shape in the three dimensional Heisenberg magnet is essentially
influenced by a surface anisotropy. We predict that depending of the surface
anisotropy type there appears barrel-- or pillow--shaped deformation of the vortex
core along the magnet thickness. Our theoretical study is well confirmed by spin-lattice simulations.
\end{abstract}




\maketitle

\section{Introduction}
\label{sec:intro}

Among different nontrivial magnetization distributions in the nanoscale, magnetic vortices attract a special interest because the vortex configuration can form a ground state in nano- and micron-sized ferromagnets. It takes place when the sample size exceeds the single-domain size due to the competition between exchange field and a stray one in magnets with small magnetocrystalline anisotropy \cite{Hubert98,Guimaraes09}. Nontrivial topological properties of vortices \cite{Braun12} attract interest to their study with perspective application to the high-density magnetic storage devices, nonvolatile magnetic vortex random-access memories \cite{Guimaraes09,Yu11a}.

In common with stray field effects which favour the vortex configuration, the vortex can form the lowest energy state in magnets with a surface anisotropy \cite{Kireev03,Leonel07}. Such anisotropy, which always appears in real samples, is originated from the symmetry breaking for the boundary sites of the lattice and can result in the specific uniaxial single-ion anisotropy of different sings \cite{Kireev03,Fiorani05}. In the disk-shaped magnets the edge surface anisotropy can pin the magnetization along the border in the circular, i.e. in the vortex, configuration \cite{Kireev03}. Similarity between the effects of the stray field and the surface anisotropy is not casual: Effective surface anisotropy in thin nanomagnets is known to be induced by the dipolar interaction \cite{Kohn05a,Caputo07b}.

In this work we study analytically and numerically the influence of the single-ion uniaxial surface anisotropy of different types, easy--surface (ES) and easy--normal (EN), on the three-dimensional (3D) vortex shape for the Heisenberg magnet. We show that the presence of the surface anisotropy breaks the symmetry of magnetization structure in the axial $\hat{\vec{z}}$--direction, which naturally leads to $\hat{\vec{z}}$--dependence of the vortex core width: there appears the barrel- and the pillow- deformation of the core for the ES and EN anisotropies, respectively.

\section{The model}
\label{sec:model}

The model we consider is a ferromagnetic system, described by the classical Heisenberg Hamiltonian
\begin{subequations} \label{eq:H-total}
\begin{equation} \label{eq:H-ex}
\mathcal H =  - J\mathcal{S}^2 \sum_{(\vec n, \vec \delta)} \vec m_{\vec n} \cdot \vec m_{\vec n + \vec \delta} + \mathcal{H}^{\text{an}},
\end{equation}
where $J>0$ is the exchange integral, $\mathcal S$ is the length of classical spin, $\vec{m}_{\vec{n}}$ is the normalized magnetic moment on a 3D site position $\vec{n}$, the 3D index $\vec \delta$ runs over the nearest neighbours, and $\mathcal{H}^{\text{an}}$ is the anisotropy part of the Hamiltonian. We take into account the bulk on--site anisotropy with the constant $K>0$ (easy--plane anisotropy) and the surface one with the surface anisotropy constant $K_s$ \cite{Neel54,Fiorani05}
\begin{equation} \label{eq:H-EP+SA}
\mathcal{H}^{\text{an}} = \frac{K\mathcal{S}^2}{2}\sum_{\vec n} (\vec m_{\vec n} \cdot \hat{\vec{z}})^2  - \frac{K_s\mathcal{S}^2}{2} \sum_{(\vec{l}, \vec{l'})} (\vec m_{\vec l} \cdot \vec{u}_{\vec{l}\vec{l'}})^2.
\end{equation}
\end{subequations}
Here the last term describes the N{\'e}el surface anisotropy with the unit vector $\vec{u}_{\vec{l}\vec{l'}}$ connecting the magnetic moment $\vec{m}_{\vec{l}}$ from the surface site $\vec{l}$ to its nearest neighbour $\vec{l'}$.

The continuum description of the system is based on smoothing the lattice model using the normalized magnetization
\begin{equation} \label{eq:m}
\begin{split}
\vec{m}(\vec{r},t) &= a^3\sum_{\vec{n}} \vec{m}_{\vec{n}} \delta\left(\vec{r} - \vec{r}_{\vec{n}} \right)\\
&= \left(\sin\theta \cos\phi, \sin\theta \sin\phi, \cos\theta \right),
\end{split}
\end{equation}
where $\theta = \theta(\vec r,t)$, $\phi = \phi(\vec r,t)$, the parameter $a$ being the lattice constant, and $\delta(\vec{r})$ being the Dirac $\delta$--function.

The total energy, the continuum version of the Hamiltonian \eqref{eq:H-total}, normalized by $K\mathcal{S}^2/a^3$ has the following form
\begin{equation} \label{eq:Etot}
\begin{split}
\mathscr{E} &\equiv \frac{E}{K\mathcal{S}^2/a^3} = \mathscr{E}_v + \mathscr{E}_s,\\
\mathscr{E}_v &=\frac{1}{2}\int \mathrm{d}V \left[-\ell^2 \vec{m}\cdot\vec{\nabla}^2\vec{m} + (\vec{m}\cdot \hat{\vec{z}})^2\right],\\
\mathscr{E}_s &= \frac{\varkappa a}{2}\int \mathrm{d}S (\vec{m}\cdot \vec{n}_{s})^2
\end{split}
\end{equation}
with $\ell = a \sqrt{J/K}$ being the magnetic length. The last term $\mathscr{E}_s$ is the transverse surface anisotropy, the continuum analogue of N{\'e}el surface anisotropy with $\vec{n}_s$ being the normal to the surface and the parameter $\varkappa = K_s/K$ being the surface anisotropy rate. In the further study we consider the cases of both ES anisotropy when $\varkappa>0$ and the EN one when $\varkappa<0$.

The equilibrium magnetization structure can be found by variation of the energy functional \eqref{eq:Etot}, which results in the following boundary-value problem: \cite{Brown63,Hubert98}
\begin{subequations} \label{eq:BVP-tot}
\begin{align} \label{eq:BVP-tot-1}
\vec m \times \left[ \ell^2\vec\nabla^2\vec{m} - (\vec{m}\cdot \hat{\vec{z}}) \hat{\vec{z}} \right] = 0,\\
\label{eq:BVP-tot-2}
\ell^2\frac{\partial \vec{m}}{\partial \vec{n}_s}\Biggr|_{S} = \varkappa a (\vec{m}\cdot\vec{n}_s)  \left[ (\vec{m}\cdot\vec{n}_s) \vec{m} - \vec{n}_s\right]\Biggr|_{S}.
\end{align}
\end{subequations}
One can see that the presence of the surface anisotropy changes the symmetry of boundary conditions, leading to the Robin boundary conditions instead of the Neumann ones \cite{Weisstein98}. As a result the symmetry breaking the magnetization structure becomes $\hat{z}$--dependent. In particular, we will see that the vortex core width takes different values in a volume and on the surface.

\section{Vortex core structure: analytics}
\label{sec:analytics}

Let us consider the disk--shape sample the radius $R$ and the thickness $L$. The volume contribution to the energy functional \eqref{eq:Etot} reads
\begin{subequations} \label{eq:Etot-disk}
\begin{equation} \label{eq:Etot-disk-vol}
\mathscr{E}_v  = \frac{1}{2}\int \mathrm{d}V \left\{\ell^2\left[(\vec\nabla \theta)^2 + \sin^2\theta (\vec\nabla \phi)^2\right]+\cos^2\theta\right\},
\end{equation}
where $\hat{\vec{z}}$ is the cylinder axis. The surface energy term $\mathscr{E}_s = \mathscr{E}_s^{\mathrm{face (+)}} + \mathscr{E}_s^{\mathrm{face (-)}} + \mathscr{E}_s^{\mathrm{edge}}$,
\begin{equation} \label{eq:Etot-disk-surf}
\begin{split}
\mathscr{E}_s^{\mathrm{face (\pm)}}&= \frac{\varkappa a}{2}\int \mathrm{d}S^{\mathrm{face (\pm)}} \cos^2\theta \Biggr|_{z=\pm L/2} ,\\
\mathscr{E}_s^{\mathrm{edge}} &= \frac{\varkappa a}{2}\int \mathrm{d}S^{\mathrm{edge}} \sin^2\theta  \cos^2(\phi-\chi) \Biggr|_{\rho=R},
\end{split}
\end{equation}
\end{subequations}
where $(\rho,\chi,z)$ are the cylinder coordinates.

In terms of the angular variables the boundary--value problem \eqref{eq:BVP-tot} for the disk--shaped sample has the following form:
\begin{subequations} \label{eq:BVP-disk}
\begin{align} \label{eq:BVP-disk-1}
&\vec\nabla^2\theta - \frac{1}{2}\sin2\theta \left[(\vec\nabla\phi)^2 - \frac{1}{\ell^2}\right]=0,\\
\label{eq:BVP-disk-2} %
&\vec\nabla\cdot\left(\sin^2\theta\vec\nabla\phi \right)=0,\\
\label{eq:BVP-disk-3} %
&\pm \ell^2 \frac{\partial \theta}{\partial z} - \frac{\varkappa a}{2}  \sin2\theta \Biggr|_{z=\pm L/2} \!\!\!\!\! = 0,\quad
\frac{\partial \phi}{\partial z} \Biggr|_{z=\pm L/2} \!\!\!\!\! = 0,\\
\label{eq:BVP-disk-4} %
&\ell^2 \frac{\partial \theta}{\partial \rho} + \frac{\varkappa a}{2}  \sin2\theta\cos^2(\phi-\chi) \Biggr|_{\rho=R} \!\!\!\!\! = 0,\\
\label{eq:BVP-disk-5} %
&\ell^2 \frac{\partial \phi}{\partial \rho} - \frac{\varkappa a}{2} \sin2(\phi-\chi)\Biggr|_{\rho=R} \!\!\!\!\! = 0.
\end{align}
\end{subequations}

The form of boundary conditions determines possible minimizers. One can see that the boundary--value problem \eqref{eq:BVP-disk} has the vortex--like stationary solution with
\begin{subequations} \label{eq:vortex}
\begin{equation} \label{eq:planar-vortex}
\phi = \chi+\varphi_0.
\end{equation}
To satisfy the boundary condition \eqref{eq:BVP-disk-5}, the value of the constant $\varphi_0=\pm\pi/2$ for $\varkappa>0$ (ES magnets) and $\varphi_0=0,\pi$ for $\varkappa<0$ (EN magnets).

The simplified version of the boundary--value problem \eqref{eq:BVP-disk} with $\theta=\pi/2$ was considered in Refs.~\cite{Kireev03,Moser04,Kurzke06}: Planar vortices with $\cos\theta=0$ and $\phi=\chi+\varphi_0$ were shown to be metastable states in the disk--shaped system.

Below we discuss the 3D boundary--value problem \eqref{eq:BVP-disk}. In this case the nonplanar vortex with $z$--dependence of the polar angle appear:
\begin{equation} \label{eq:nonplanar-vortex}
\theta = \theta(\rho,z).
\end{equation}
\end{subequations}
The typical scale of the $\theta$--distribution is determined by the magnetic length $\ell$. Supposing that $\ell\ll R$, we can replace the boundary condition \eqref{eq:BVP-disk-4} by
\begin{equation} \label{eq:theta-inf}
\frac{\partial \theta}{\partial \rho}\Biggr|_{\rho=R\to\infty}\!\!\!\! =0, \qquad \cos\theta\Biggr|_{\rho=R\to\infty} \!\!\!\!\! = 0.
\end{equation}

The problem \eqref{eq:BVP-disk} is the nonlinear boundary--value problem for the partial differential equation for the function \eqref{eq:nonplanar-vortex}. To simplify the analysis we use the variational approach with Ansatz--function
\begin{equation} \label{eq:vortex-Ansatz}
\cos\theta(\rho,z) = f\left(\frac{\rho}{w(z)\ell}\right), \qquad f(x) = \exp\left(-\frac{x^2}{2}\right).
\end{equation}
This function is the generalization of the well--known Feldt\-keller Ansatz \cite{Hubert98,Feldtkeller65}, originally used to describe the structure of the vortex in thin films. However in contrast to \cite{Hubert98} our reduced vortex core function $w(z)$  is a variational function.

\begin{figure*}
\begin{subfigure}{0.25\textwidth}
\includegraphics[width=\columnwidth]{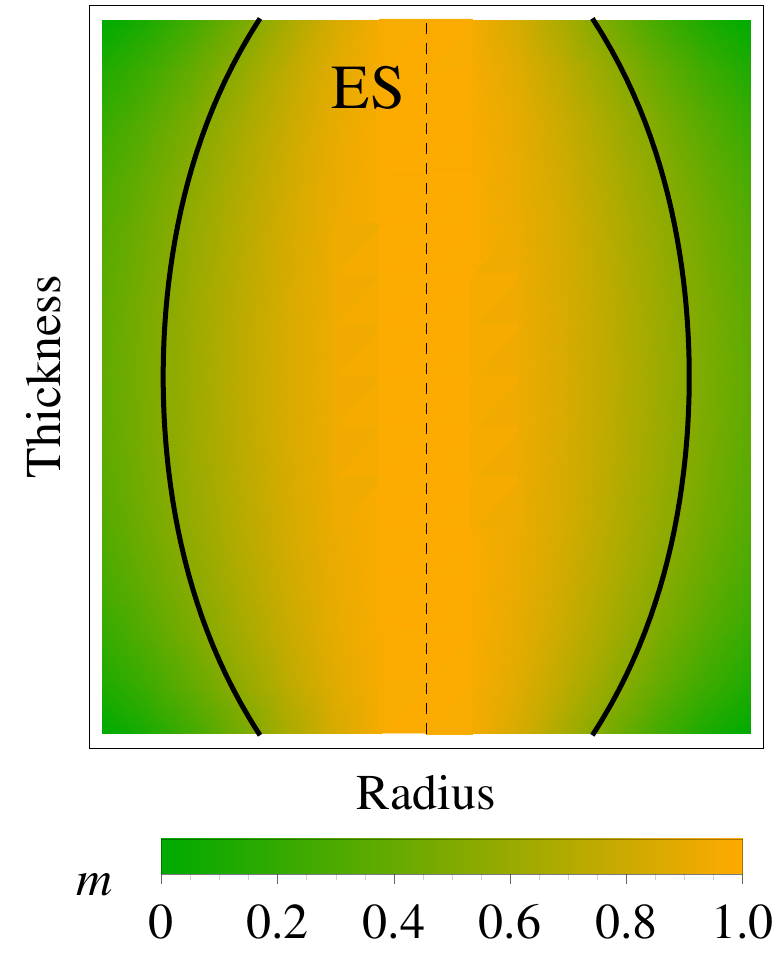}
\caption{Schematic of the barrel--shaped vortex width in case of easy--surface surface anisotropy}
\end{subfigure}
\begin{subfigure}{0.49\textwidth}
\includegraphics[width=\columnwidth]{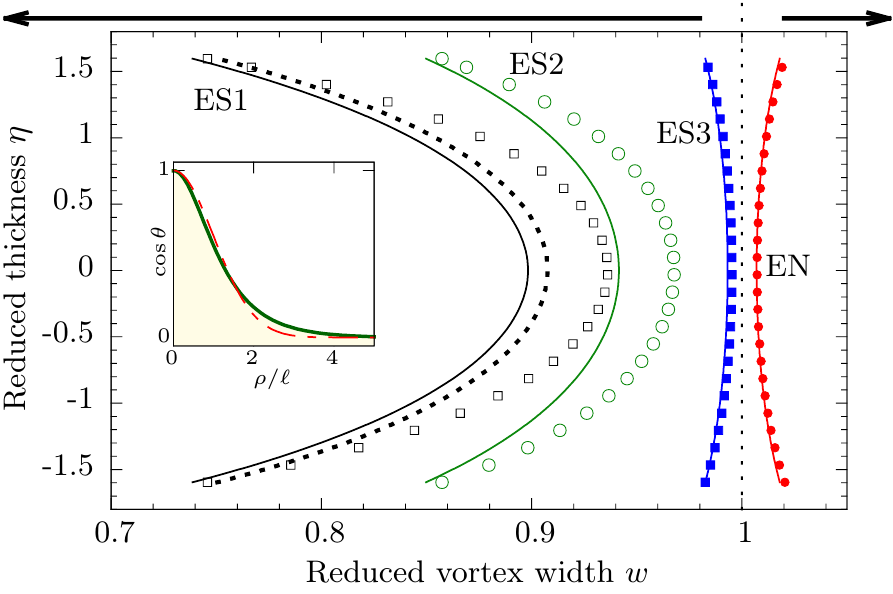}
\caption{Comparison of the analytical result and simulations}
\end{subfigure}
\begin{subfigure}{0.25\textwidth}
\includegraphics[width=\columnwidth]{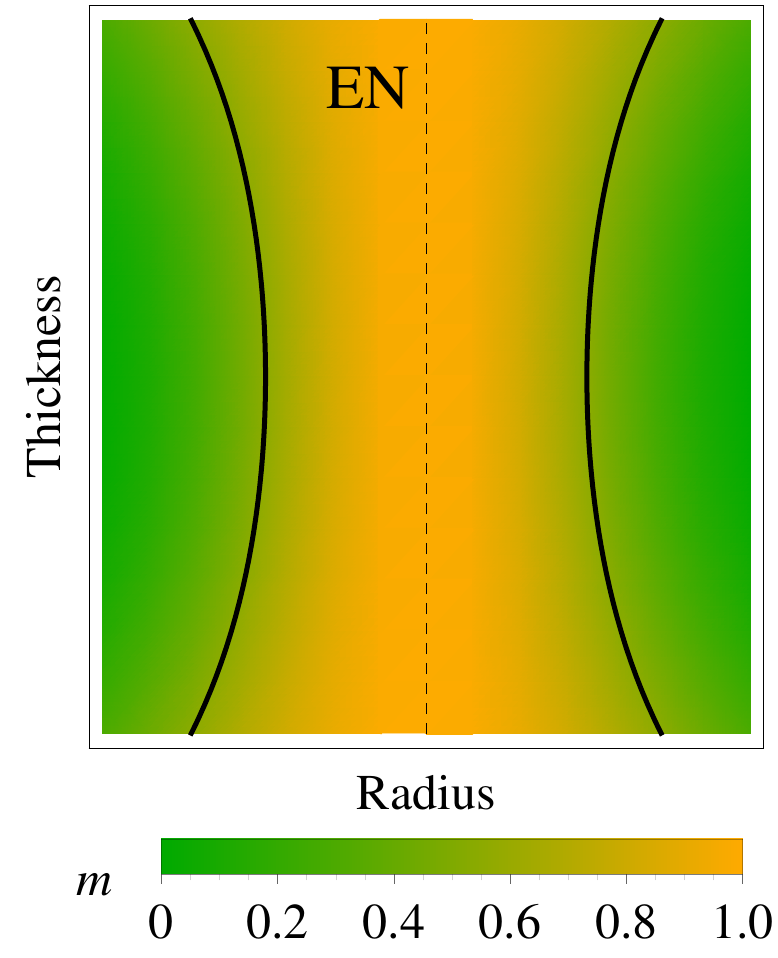}
\caption{Schematic of the pillow--shaped vortex width in case of easy--normal surface anisotropy}
\end{subfigure}

\caption{(Colour online) \textit{Centre.} Vortex core width as a function of sample thickness coordinate for different anisotropies: easy--surface with curves ES1 ($\varkappa = 10$), ES2 ($\varkappa = 5$), ES3 ($\varkappa = 0.5$) and easy--normal EN ($\varkappa = -0.5$). Symbols correspond to the simulations and solid lines to the analytical dependence~\eqref{eq:w-linearized}. Dashed curve for ES1 corresponds to the numerical solution of the Eq.~\eqref{eq:w-BVP}. Parameters: $\ell = 14a$, $L=49a$, $2R=249a$. Thin dotted line $w(z)=1$ shows the equilibrium value for absence of surface anisotropy. The ``easy--surface'' curves lie left and ``easy--normal'' ones lie right of it. Inset shows the central part of the out-of-plane vortex shape from the simulations (solid curve) and by Ansatz~\eqref{eq:vortex} (dashed-dot line) in the absence of surface anisotropy. \textit{Left and right.} Schematic of the vortex core shape for the easy--surface and easy--normal surface anisotropies respectively. Colour shows the 
magnitude 
of $m$ and
solid lines show the thickness of the vortex. Dashed lines show the centre of the sample $r=0$. }
\label{fig:w-via-z}
\end{figure*}

Using Ansatz \eqref{eq:vortex-Ansatz} one can write down the energy in the form
$\mathscr{E} = \mathscr{E}_0 + \pi\ell^3\sqrt{\zeta(3)} \widetilde{\mathscr{E}}[w]$, where the first term $\mathscr{E}_0$ is independent of the $z$ coordinate, the second term $\widetilde{\mathscr{E}}[w]$ contains terms both due to the volume contribution as due to the surface anisotropy, and $\zeta(3)$ is the Ap\'{e}ry's constant, see \ref{sec:appendix} for details
\begin{equation} \label{eq:energy-func}
\begin{split}
\widetilde{\mathscr{E}}[w] &= \!\! \int_{-\lambda}^{\lambda} \!\! \mathrm{d} \eta \left[ w'^2(\eta) -\ln w(\eta) + \frac{w^2(\eta)}{2}\right]\\
&+\frac{\tilde\varkappa}{2} \left[w^2(\lambda)+w^2(-\lambda) \right] , \qquad \lambda = \frac{L}{2\ell\sqrt{\zeta(3)}}.
\end{split}
\end{equation}
Here the prime denotes the derivative with respect to dimensionless thickness coordinate, $\eta = z/[\ell\sqrt{\zeta(3)}]$, see \eqref{eq:t-n-lambda} and $\tilde\varkappa = \varkappa a /[\ell \sqrt{\zeta(3)}]$.

Now we are able to calculate the equilibrium vortex width varying the functional \eqref{eq:energy-func}, which results
in the following boundary-value problem:
\begin{subequations} \label{eq:w-BVP}
\begin{align} \label{eq:w-eq}
&2 w''(\eta) =  w(\eta) - \frac{1}{w(\eta)}, \\
\label{eq:w-BC} %
&\tilde\varkappa w(\eta) \pm 2w'(\eta)\Bigr|_{\eta=\pm \lambda} = 0.
\end{align}
\end{subequations}
Note that without the surface anisotropy \eqref{eq:w-BC} takes the form of Neumann boundary conditions, $w'(\eta)\bigr|_{\eta=\pm \lambda} = 0$, hence the problem \eqref{eq:w-BVP} has the only constant solution $w(\eta)=1$, which is the reduced core width for $\varkappa=0$.

One can easily analyze \eqref{eq:w-BVP} in the case of weak surface anisotropy, when $|\tilde\varkappa|\ll1$. In this case the value of normalized vortex width $w$ is close to unity, hence $w(\eta) = 1 - x(\eta)$, where $|x(\eta)| \ll 1$. Below we verify this assumption by comparison with simulations. Now by linearizing the Eq.~\eqref{eq:w-eq} with respect to $x$ with account of Robin boundary conditions \eqref{eq:w-BC}, one can calculate $w(\eta)$
\begin{equation} \label{eq:w-linearized}
w(\eta) = 1 - \frac{\tilde\varkappa  \cosh \eta}{\tilde \varkappa  \cosh \lambda + 2 \sinh \lambda}.
\end{equation}
One can see that due to the surface anisotropy the vortex core width varies with $z$ coordinate and its shape is dependent on the sign of $\tilde\varkappa$, i.e. the type of the surface anisotropy.

Typical results for the $z$--dependence of the vortex core width, $w(\eta)$ are shown in the Fig.~\ref{fig:w-via-z}: the vortex core becomes barrel--shaped for ES surface anisotropy ($\varkappa>0$) and the pillow--shaped for EN one ($\varkappa<0$). Numerically calculated vortex core profile as solution of \eqref{eq:w-BVP} is plotted by solid curves for different values of surface anisotropy constants. One can see that the analytically calculated profile (dashed curve ES1) agrees well with numerical calculations.

\section{Vortex core structure: spin--lattice simulations}
\label{sec:SLaSi}

To verify our predictions about the 3D shape of the vortex numerically we simulate a 3D magnet with a simple cubic lattice and the surface anisotropy using in-house developed spin-lattice simulator \textsf{SLaSi} \cite{SLaSi}. The system is described by the discrete Landau--Lifshitz--Gilbert equations for the Hamiltonian \eqref{eq:H-total}
\begin{equation} \label{eq:LLG}
\frac{\mathrm d\vec m_{\vec n}}{\mathrm d\tau} = \vec m_{\vec n} \times \frac{\partial \mathscr H}{\partial \vec m_{\vec n}} + \varepsilon \vec m_{\vec n} \times \frac{\mathrm d\vec m_{\vec n}}{\mathrm d\tau},
\end{equation}
where $\tau = K\mathcal{S}t/\hslash$ is the dimensionless time, $\mathscr  H = \mathcal H a^3/(K\mathcal{S}^2)$ is the dimensionless energy, $\hslash$ is the Plank's constant, $\varepsilon$ is the Gilbert damping, and $\vec{n}$ the 3D index running over spin lattice. \footnote{Integration is performed by the modified fourth- and fifth-order Runge--Kutta--Fehlberg method (RKF45) and free spins on the surface of the sample, see Ref.~\cite{SLaSi} for details.}

We model the vortex distribution in such 3D spin lattice without external fields. We consider the disk sample with thickness of 50 sites, $L=49a$, diameter $2R=249a$. Since we are interested in static structure, we consider the overdamped regime by choosing the Gilbert damping constant $\varepsilon = 0.5$. We use $\ell=14a$ for detailed description of the vortex core.

The typical vortex structure obtained from simulations is presented in Fig.~\ref{fig:w-via-z}. The inset in the centre panel shows the out-of-plane vortex shape $\cos\theta$ as a function of dimensionless radius $\rho/l$ for the case $\varkappa=0$. One can see that the Ansatz function \eqref{eq:vortex-Ansatz} (dashed-dot line) provides a close approximation to the simulation data (solid curve).

In the presence of the surface anisotropy the vortex structure is changed in accordance to our theory described above. We perormed our simulations for different values of surface anisotropy. Here we present results for the values ES surface anisotropy with $\varkappa_1 = 0.5$, $\varkappa_2 = 5$ and $\varkappa_3 = 10$, and one value of EN surface anisotropy $\varkappa_4 = -0.5$. The smaller values of $|\varkappa|$ are more realistic, nevertheless the strong surface anisotropy allows to obtain the sharp effect for comparison with theory. Note that the EN surface anisotropy coefficient has to be smaller that the volume coefficient in order to provide an effective total easy--plane effective anisotropy which supports the vortex state.  Symbols in the Fig.~\ref{fig:w-via-z} are obtained by fitting the simulated vortex structure with the use of Eq.~\eqref{eq:vortex-Ansatz}.

The vortex core width of the relaxed vortex varies with the thickness coordinate. The barrel--shaped or the pillow--shaped form of the vortex core depend on the surface anisotropy type. In the case of the easy--surface surface anisotropy the total anisotropy coefficient per site $K+|K_s|$ on the face surface sites is larger than inside volume ones. It results in the decreasing of the effective magnetic length
\begin{equation}\label{eq:l-eff}
\ell_\text{eff} =  a \sqrt{\frac{J}{K+\left|K_s\right|}}
\end{equation}
on the face surfaces. Due to the exchange coupling between lattice layers a smooth change of the $\ell_\text{eff}$ occurs. It reaches the maximal value in the centre of the sample's axis and one observes the barrel--shaped vortex profile.

The comparison of the Eq.~\eqref{eq:w-linearized} with simulations is shown on the Fig.~\ref{fig:w-via-z} by solid lines. The equilibrium value of the reduced vortex width in the absence of the surface anisotropy is $w=1$. The presence of the surface anisotropy shifts it to be thinner for the easy--surface surface anisotropy and wider for the easy--normal one. The inset shows the comparison of the vortex shape from simulations with the Ansatz~\eqref{eq:vortex}.

\section{Discussion}
\label{sec:discussion}

Let us discuss how the predicted effects can influence the vortex statics and dynamics in nanomagnetic particles.
Typical magnetic nanodots are the samples of confined ferromagnet with weak anisotropy. The vortex configuration in such magnets can be their ground state. For example, in Permalloy disk-shaped samples the vortex state is realized if the disk diameter exceeds 50 nm\cite{Guimaraes09}. The reason is that the vortex configuration favours the dipolar interaction (the stray field is absent for the pure vortex state). This can be explained by the language of effective anisotropy, induced by the dipolar interaction \cite{Caputo07b}. For thin magnets the effective spatial-dependent anisotropy contribution is localized near the edge of the particle so that magnetization will be tangent to the boundary \cite{Caputo07b}. This confirms the notion of a surface edge anisotropy.

The vortex core creates nonvanishing surface magnetostatic charges of opposite signs on the different face surfaces. Their minimization leads to a complicated vortex profile both in plane and in the axial direction~\cite{Hubert98}. When the nanodot thickness becomes comparable with a few exchange lengths, the thickness of the Bloch line appreciably varies along the thickness coordinate, see~\cite{Hubert98,Usov08a}. This conclusion agrees with our statement about the varying width $w(z)$ defined in Eq.~\eqref{eq:vortex}.

In this context it is instructive to make a link with another way to describe the inhomogeneity of the vortex width proposed in Ref.~\cite{Hubert98}. The trial functions $ m = \sum_i c_i \exp\left[ -2r^2/\ell_i^2w_i(z) \right]$ and $w_i(z) = b_i(1 - 4z^2/L^2)$ was used to describe an influence of magnetostatics, where $\sum_i c_i = 1$ and the radial component of magnetization is obtained from the condition of the absence of the total volume magnetostatic charge. The unknown coefficients $b_i$, $c_i$ and $\ell_i$ were calculated through minimization of the total energy. In the our surface anisotropy approach the expression for $w$ can be obtained directly with enough accuracy.

In conclusion, the shape of the vortex in the 3D Heisenberg magnet with single-ion bulk and surface anisotropies is studied analytically and numerically. It is shown that the vortex width varies along the disk axis and its shape is dependent on the sign of the surface anisotropy coefficient. Vortex becomes barrel--shaped for the easy--surface surface anisotropy and pillow--shaped for the easy--normal one.

Taking into account surface anisotropy, the magnetization distribution naturally becomes inhomogeneous in the axial direction. In the effect of the axial dependence of the magnetization it is similar to the influence of the stray fields. The demagnetizing field in the vortex core region causes the twisting of the in-plane core magnetization and barrel--shaped vortex width along the nanodisk axis~\cite{Hubert98, Usov08a}. For sufficiently thin samples a homogeneous magnetization distribution along thickness is usually used~\cite{Kravchuk07a} and the polarity reversal occurs through the planar vortex formation. In the Permalloy nanodisks of thickness~50\,nm is was shown numerically that it  is accompanied by the Bloch points nucleation~\cite{Thiaville03}. We expect that the surface anisotropy for the nanoparticles in the vortex state allows to expect the change of the polarity switching mechanism to the second one.

\section*{Acknowledgements}

The authors thank Prof. F.~Mertens for helpful discussions. O.~P. and D.~S. thank the University of Bayreuth, where a part of this work was performed, for kind hospitality. O.~P. thanks Computing Center of the University of Bayreuth~\cite{btrzx}. The present work was partially supported by the Grant of President of Ukraine for support of young scientists researches (Project No GP/F49/083).

\appendix

\section{Energy calculation}
\label{sec:appendix}

Let us start from the energy functional \eqref{eq:Etot-disk}. Substituting the vortex solution in the form \eqref{eq:vortex} one can rewrite the vortex energy in the following form
\begin{equation*}
\begin{split}
\mathscr{E}_v  &= \pi\!\! \int_{-L/2}^{L/2} \!\! \mathrm{d}z \!\!\int_0^{R} \!\! \rho \mathrm{d}\rho \Biggl\{\ell^2 \left[ \left(\frac{\partial \theta}{\partial \rho}\right)^2 + \left(\frac{\partial \theta}{\partial z}\right)^2+\frac{\sin^2\theta}{\rho^2}\right]\\
& + \cos^2\theta\Biggr\}, \qquad \mathscr{E}_s^{\mathrm{face (\pm)}} = \pi \varkappa a\int_0^R \rho\mathrm{d}\rho \cos^2\theta \Biggr|_{z=\pm L/2}.
\end{split}
\end{equation*}
Now using the Ansatz \eqref{eq:vortex-Ansatz} this energy can be simplified as follows
\begin{equation} \label{eq:energy-func-1}
\begin{split}
\mathscr{E} &= \mathscr{E}_0 + \mathscr{E}_1[w],\\
\mathscr{E}_0 &=\pi \ell^2 L \left(\ln \frac{R}{\ell} + \frac{\pi^2}{12} + \frac{\gamma}{2}\right),\\
\mathscr{E}_1[w] &= \pi \ell^2\int_{-L/2}^{L/2} \mathrm{d} z \Biggl[ \ell^2\zeta(3) \left(\frac{\mathrm{d} w}{\mathrm{d} z} \right)^2 -\ln w + \frac{w^2}{2}\Biggr]\\
& + \frac{\pi \varkappa a\ell^2}{2}w^2\Biggr|_{z=\pm L/2},
\end{split}
\end{equation}
where  $\zeta(3)\approx 1.202$ is the Ap\'{e}ry's constant and $\gamma \approx 0.5772$ is the Euler's constant \cite{Weisstein98}. Now by changing variables
\begin{equation} \label{eq:t-n-lambda}
\eta = \frac{z}{\ell\sqrt{\zeta(3)}}, \qquad \lambda = \frac{L}{2\ell\sqrt{\zeta(3)}},
\end{equation}
one can rewrite \eqref{eq:energy-func-1} in the form \eqref{eq:energy-func}.

%
%

\end{document}